\documentclass[a4paper,english,reprint, aps]{revtex4}
\usepackage[T1]{fontenc}
\usepackage[latin9]{inputenc}
\usepackage{mathtools}
\usepackage{amsmath}
\usepackage{amsthm}
\usepackage{amssymb}

\makeatletter

\pdfpageheight\paperheight
\pdfpagewidth\paperwidth

\@ifundefined{textcolor}{}
{%
 \definecolor{BLACK}{gray}{0}
 \definecolor{WHITE}{gray}{1}
 \definecolor{RED}{rgb}{1,0,0}
 \definecolor{GREEN}{rgb}{0,1,0}
 \definecolor{BLUE}{rgb}{0,0,1}
 \definecolor{CYAN}{cmyk}{1,0,0,0}
 \definecolor{MAGENTA}{cmyk}{0,1,0,0}
 \definecolor{YELLOW}{cmyk}{0,0,1,0}
}
\theoremstyle{plain}
\newtheorem{thm}{\protect\theoremname}
  \theoremstyle{plain}
  \newtheorem{ax}[thm]{\protect\axiomname}

\makeatother

\usepackage{babel}
  \providecommand{\axiomname}{Axiom}
\providecommand{\theoremname}{Theorem}

\begin{document}

\preprint{APS/123-QED}

\title{Axiomatic Local Metric Derivatives for Low-Level Fractionality with
Mittag-Leffler Eigenfunctions}

\thanks{The authors wish to express their gratitude to FAPERJ-Rio de Janeiro
and CNPq-Brazil for the partial financial support.}

\author{J. Weberszpil}

\email{josewebe@gmail.com}

\affiliation{Universidade Federal Rural do Rio de Janeiro, UFRRJ-IM/DTL\\
 Av. Governador Roberto Silveira s/n- Nova Iguaçú, Rio de Janeiro,
Brasil, 695014.}

\author{J. A. Helayël-Neto}

\email{helayel@cbpf.br}

\address{Centro Brasileiro de Pesquisas Físicas-CBPF-Rua Dr Xavier Sigaud
150,\\
 22290-180, Rio de Janeiro RJ Brasil. }

\date{\today}
\begin{abstract}
In this contribution, we build up an axiomatic local metric derivative
that exhibits Mittag-Leffler function as an eigenfunction and is valid
for low-level fractionality, whenever the order parameter is close
to $1$. This version of deformed (or metric) derivative may be a possible
alternative to the versions worked out by Jumarie and the so-called
local fractional derivative also based on Jumarie's approach.
With rules similar to the classical ones, but with a systematic axiomatic
basis in the limit pointed out here, we present our results and some
comments on the limits of validity for the controversial formalism
found in the literature of the area.

\medskip{}

\end{abstract}


\maketitle

\section{Introduction}

In recent works, we have developed connections and a variational formalism
for deformed or metric derivatives, considering the relevant space-time/
phase space as fractal or multifractal \citep{Nosso On conection2015,Variational Deformed}.

The use of deformed-operators was justified based on our proposition
that there exists an intimate relationship between dissipation, coarse-grained
media and the some limit scale of energy for the interactions. Concepts
and connections like open systems, quasi-particles, energy scale and
the change of the geometry of space\textendash time at its topology
level, nonconservative systems, noninteger dimensions of space\textendash time
connected to a coarse-grained medium, all of this were discussed.
With this perspective, we argued that deformed or we should say, metric
derivatives, similarly to the Fractional Calculus (FC), could allows
us to describe and emulate certain dynamics without explicit many-body,
dissipation or geometrical terms in the dynamical governing equations.
Also, we emphasized that the paradigm we adopt were different from
the standard approach in the generalized statistical mechanics context
\citep{Tsallis1,Tsallis BJP- 20 anos,Tsallis2}, where the modification
of entropy definition leads to the modification of algebra and consequently
the derivative concept \citep{Nosso On conection2015,Variational Deformed}.
This was structured by mapping to a continuous fractal space \citep{Balankin PRE85-Map-2012,Balankin Rapid Comm,Balankin-Towards a physics on fractals}
which leads naturally to the necessity of modifications in the derivatives,
that we called deformed or better than, metric derivatives \citep{Nosso On conection2015,Variational Deformed}.
The modifications of derivatives, accordingly with the metric, brings
to a change in the algebra involved, which in turn may conduct to
a generalized statistical mechanics with some adequate definition
of entropy.

In this communication, we develop, by means of an axiomatic approach,
a local metric derivative that has the Mittag-Leffler function as
an eigenfunction and for which deformed Leibniz' and the chain rule hold, similarly
to the standard calculus, but whenever a low-level of fractionality is considered.
The deformed operators here are local.

Our main goal here is to show that it is possible to build up a local
formalism that meets certain similarities to those required by a version
of fractional calculus attributed to Jumarie \citep{Jumarie1,Jumarie2-Table,Jumarie 2013},
in such a way that reconciles the relevant axioms of the approach
outlined here and, additionally, it is an approximation valid within
a low fracionality limit. In this way, we try to make clear, in short,
that existing studies in the literature that apply the Modified Riemann
Liouville (MRL) derivative definition \citep{Jumarie1}, and there
are so many, are in fact local and behave as an approximations but,
provided that the order of the derivative is very close to one. That
is, being local and considered as approximations.

Our paper is outlined as follows: In Section 2, some motivations are
collected; in Section 3, we cast some mathematical aspects; in Section
4, we build up an axiomatic local metric derivative. We finally present
our Conclusions and Outlook in Section 5.

\section{The motivation}

Some initial claims here coincide with our work of Refs. \citep{Nosso On conection2015,Variational Deformed}.
The local differential equation $\frac{dy}{dx}=y^{q},$ yields the
solution given by the q-exponential, $y=e_{q}^{x}$ \citep{Tsallis1,Tsallis BJP- 20 anos,Tsallis2}.

The differential equation, with the local Hausdorff derivative proposed
in Ref. \citep{Chen-Time Fabic}, reads:

\begin{equation}
\frac{d^{H}y}{dx^{\alpha}}=y,
\end{equation}
and leads to the stretched exponential solution $y=e^{x^{\alpha}}$.

But, in terms of nonlocal fractional calculus, the fractional differential
equation 
\begin{equation}
\frac{d^{\alpha}y}{dx^{\alpha}}=y,
\end{equation}
with the Caputo fractional derivative approach, yields the solution
in terms of the Mittag-Leffler function $y=E_{\alpha}(x^{\alpha})$.

The Mittag-Leffler function is of extreme importance to describe the
dynamics of complex systems. It involves a generalization of exponential
function and several trigonometric ands hyperbolic functions.

The problem with FC is that Leibniz rule and the chain rule of traditional
calculus are violated in their standard forms, which makes it difficult
to apply the mathematical tool in areas such as field and particle
physics. Other problems are the fact that the derivative of a constant
is not zero in the Riemann-Liouville approach and that there are strong
restrictions for the class of smooth functions in the Caputos's approach.

Some authors have been attempting to come over those difficulties
by redefining some expressions of fractional integral and derivatives
\citep{Jumarie1}, arguing that by this way the basic rules of usual
differential and integral calculus would be non-violated. Unfortunately
several mistakes here pointed out for this attempts \citep{Tarasov-No violation,Tarasov On Chain,Chines critica Jumarie,critica Jumarie 2,Tenreiro-What}.
Recently, we have suggested that the mentioned approaches were in
fact based on local operators \citep{Nosso On conection2015}.

In this sense, there emerges a question: is it possible to have a
local deformed operator whose eigenfunctions could be the Mittag-Leffler
function and preserves the non-violated form of Leibniz rule, with
some easy form for chain rule, even if the rules were considered as
approximations? This is our actual goal here. To propose some local
deformed or metric derivatives that should satisfy those requests,
but being local and an approximation for low level of fractionality
(order parameter close to $1$).

An important point to consider is a possible application for the approach
proposed here. These considerations shall be tackled in our Conclusions.

\section{A brief glance at mathematical aspects}

Here, in this Section, we provide some brief information to remember
the main forms of deformed or metric derivative. The readers may see
ref. \citep{Nosso On conection2015,Variational Deformed,Balankin PRE85-Map-2012}
for more details.

\textbf{Hausdorff Derivative}

By employing the local fractional differential operators in connection
with the Hausdorff derivative \citep{Chen-Time Fabic}, we can write
that: 
\begin{eqnarray}
\frac{d^{H}}{dx^{\zeta}}f(x) & = & \lim_{x\rightarrow x'}\frac{f(x')-f(x)}{(x')^{\zeta}-x^{\zeta}}=\nonumber \\
 & = & \left(\frac{x}{l_{0}}+1\right)^{1-\zeta}\frac{d}{dx}f=\frac{l_{0}^{\zeta-1}}{c_{1}}\frac{d}{dx}f=\frac{d}{d^{\zeta}x}f,\label{eq:Haudorff derivative}
\end{eqnarray}
where $l_{0}$ is the lower cutoff along the Cartesian $x-$axis and
the scaling exponent, $\zeta$ $,$ characterizes the density of states
along the direction of the normal to the intersection of the fractal
continuum with the plane, as defined in the work \citep{Balankin PRE85-Map-2012}.

\textbf{Conformable derivative}

This kind of deformed derivative has been proposed by the authors
of Ref. \citep{new definition}; it preserves classical operational
properties and is given by

\begin{equation}
T_{\alpha}f(t)=\lim_{\epsilon\rightarrow0}\frac{f(t+\epsilon t^{1-\alpha})-f(t)}{\epsilon}.\label{basic}
\end{equation}

If the function is differentiable in a classical sense, the definition
above yields 
\begin{equation}
T_{\alpha}f(t)=t^{1-\alpha}\frac{df(t)}{dt}.\label{eq:Differentiable}
\end{equation}

Performing the change of variable $t\rightarrow1+\frac{x}{l_{0}}$, we have shown
that (\ref{eq:Differentiable}) is nothing but the Hausdorff derivative
up to a constant and valid for differentiable functions \citep{Nosso On conection2015}.

Another similar definition of local deformed derivative with classical
properties is the one used in Ref. \citep{Aplica Katugampola}:

Let $f:[0,\infty)\rightarrow\mathbb{R}$ and $t>0$. Then, the local
deformed derivative-Katugampola- of $f$ of order $\alpha$ is defined
by, 
\begin{equation}
\mathcal{D}^{\alpha}(f)(t)=\lim_{\epsilon\rightarrow0}\frac{f(te^{\epsilon t^{-\alpha}})-f(t)}{\epsilon},\label{df}
\end{equation}
 for $t>0,\;\alpha\in(0,1)$. If $f$ is $\alpha-$differentiable
in some $(0,a),\;a>0$, and $\lim_{t\rightarrow0^{+}}\mathcal{D}^{\alpha}(f)(t)$
exists, then define
\begin{equation}
\mathcal{D}^{\alpha}(f)(0)=\lim_{t\rightarrow0^{+}}\mathcal{D}^{\alpha}(f)(t).
\end{equation}

\textbf{q-derivative in a nonextensive context}

With the generalized nonaddictive $q-$entropy as the main motivation,
the $q-$derivative sets up a deformed algebra and takes into account
that the $q-$exponential is eigenfunction of $D_{(q)}$ \citep{Borges 2004}.
Borges proposed the operator for $q$-derivative as given below:

\begin{equation}
{\displaystyle D_{(q)}f(x)\equiv{\displaystyle \lim_{y\to x}\frac{f(x)-f(y)}{x\ominus_{q}y}}}={\displaystyle [1+(1-q)x]\frac{df(x)}{dx}.}\label{eq:q-derivative}
\end{equation}

Here, $\ominus_{q}$ is the deformed difference operator, $x\ominus_{q}y\equiv\frac{x-y}{1+(1-q)y}\qquad(y\ne1/(q-1)).$

Recently, we have shown \citep{Nosso On conection2015} that the main
parameters can be connected as:

\begin{equation}
1-q=\frac{(1-\zeta)}{l_{0}}.
\end{equation}

So, we conclude that the deformed $q-$derivative is the first-order
expansion of the Hausdorff derivative and that there is a strong connection
between these formalism by means of a fractal metric.

For further more details, the reader may consult the refs. \citep{new definition,Aplica Katugampola}.

\textbf{The Leibiniz' and the chain rule hold for metric derivative}

As the reader can readily check, for the metric derivatives the Leibniz
rule holds $D^{\alpha}(fg)=gD^{\alpha}f+fD^{\alpha}g;$ and similarly,
for $q-$derivative:$D_{q}(fg)=gD_{q}f+fD_{q}g$. For composed functions,
the chain rule holds true as well:

\begin{equation}
D^{\alpha}[f\circ g](x)=\frac{df(g(x))}{dg}D^{\alpha}g(x),
\end{equation}

\begin{equation}
D_{q}[f\circ g](x)=\frac{df(g(x))}{dg}D_{q}g(x).
\end{equation}

\section{Axiomatic local metric derivative}

In this Section, we pursue the investigation of a version of metric
derivative that has the Mittag-Leffler functions as eigenfunctions.
This is due to the fact that the appearance of Mittag-Leffler function
is very present in the description of dynamics for some complex systems
like soil, porous media or some viscous systems.
\begin{ax}
Linearity
\end{ax}
$D^{\alpha}[c_{1}f(x)+c_{2}g(x)]=c_{1}D^{\alpha}f(x)+c_{2}D^{\alpha}g(x).$
\begin{ax}
Leibniz Rule
\end{ax}
$D^{\alpha}(fg)=(D^{\alpha}f)g+f(D^{\alpha}g).$
\begin{ax}
Derivative of a Power Function
\end{ax}
$D^{\alpha}(t-a)^{\nu}=\frac{\varGamma(1+\nu)}{\varGamma(1+\nu-\alpha)}(t-a)^{\nu-\alpha},$
$(\alpha>0,\,\nu>0,\,x>0).$
\begin{ax}
Low-Level Fractionality
\end{ax}
Here, we consider $\alpha$ very close to $1.$ This is what we call
low-level fractionality.

Clearly, $D_{x}^{\alpha}1=0$, since the Leibniz rule is assumed.

We consider axiom 4 to give consistency to the the simultaneous validity
of the axioms 2 and 3. So, all counter-examples for incompatibility
are actually inappropriate \citep{Tarasov-No violation,Tarasov On Chain,Tarasov Comment on Rieman}.
But, we emphasize that we are dealing here with local operators.

With axioms 3 and $D_{x}^{\alpha}1=0$, we show that the Mittag-Lefffler
functions is an eigenfunction of this local metric derivative:

$D^{\alpha}E_{\alpha}(\lambda x^{\alpha})=\lambda E_{\alpha}(\lambda x^{\alpha}).$

Of course, if we do not impose axiom 2 for Leibniz rule and substitute
the axiom by another one that attributes the derivative of a constant
to be zero, $D_{x}^{\alpha}1=0$, the result above would be valid
too. But, our intention is to permit the use o Leibniz rule, even
if it would result in an approximation.

\subsection*{From the Leibniz rule to the chain rule for local metric derivative}

We follow here a similar argument as the one presented in \citep{Tarasov-No violation,Validity},
but restricting ourselves only to local derivatives and not fractional
derivatives which are nonlocal and do not allow no-violated Leibniz
rules.

Consider a deformed or metric derivative $D^{\alpha}$ of order $\alpha$
that satisfies the Leibniz rule (axiom 2), whose domain includes all
locally Hölder-continuous functions of order $\alpha$. Using the
Hadamard's representation theorem for $f\in C^{2}$ around an arbitrary
point yields a fractional chain rule for such $D^{\alpha}$ applied
to $f\circ w$, where $w$ is locally Hölder of exponent $\alpha$,
accordingly to axiom 2. Remember axiom 4. Then the following statement
holds:

$D_{x}^{\alpha}(f\circ w)(x_{0})=f'(w(x_{0}))(D_{x}^{\alpha}w)(x_{0})$.

This kind of chain rule holds for every kind of deformed derivatives
\citep{Nosso On conection2015,new definition,Aplica Katugampola,On the local fractional derivative,Xiao-Jun Yang 1}.

\subsection*{Comments: on local fractional derivative and the Jumarie's form of
FC}

Despite the misunderstandings and missteps in the statements and formulations
of certain authors \citep{Jumarie1,Jumarie2-Table,Jumarie 2013,Xiao-Jun Yang 1},
it is worthy to highlight here some comments on possible limits of
validity and looking for a new mathematics. Given that the Jumarie's
formalism is based on the construction of a variant of FC theory,
this formalism could be seen as an approximation (in that the Mittag-Leffler
function is an eigenfunction of the fractional derivative operator
proposed, keeping certain similarities with Caputo formalism of FC,
yet, considering a low-level of fracionality limit). That is, the
Leibniz rule or the chain rule could apply to non-local fractional
operators, what actually happens not to be valid, unless we take the
low-limit fracionality\citep{Tarasov On Chain,Tarasov-No violation,Tarasov Comment on Rieman}.
However, in a recent article \citep{Nosso On conection2015}, we have
shown that the Jumarie's formalism seems to be actually related to a
local deformed derivative operator and it is not really a fractional
calculus formalism. Thus, we can consider it as an approximation,
valid for low-fracionality limit, as shown in this contribution by
means of the axiomatic form. 

Regarding the formalism called local fractional derivative \citep{Xiao-Jun Yang 1},
this controversial local version is nothing but the adaptation of
Jumarie's formalism, and therefore is only valid for low fracionality;
moreover, it carries in itself certain misconceptions of Jumarie's formalism. 

Also, taking into consideration the view of the author in Refs. \citep{Tarasov-No violation,Tarasov On Chain,Tarasov Comment on Rieman}
to deny alternatives to the usual FC formalism, our view here is based
on the attempt to look for a new mathematics that could be better
suited to the description of physical phenomena and can better model
such phenomena, especially in dealing within the context of complexity.

\section{Conclusions and outlook}

In short: we have derived an axiomatic version of metric derivative
that has the Mittag-Leffler function as an eigenfunction and is only
valid for low-level fractionality systems.

The structure of the deformed derivative generated shows similarities
with some versions claimed to hold or not hold in the literature,
but here it is axiomatically set up and the consistency are given
in the context of low level of fractionality.

We obtain a chain rule for a metric derivative of order $\alpha\simeq1$
whose domain includes $C^{1}$ or even locally Hölder-continuous functions
(which describe coarse-grained media), provided the Leibniz rule holds.

On the applicability of our approach proposed here, some comments
are deserved. Certain systems in Nature seem to be well-described
by some fractional parameter in a low-level limit of fractionality.
Examples of such low-level systems are the anomalous magnetic g-factor
of the charged leptons of the Standard Model, where the low level
of fractionality seems to be consistent with the Quantum Electrodynamics
corrections \citep{nosso AHEP-g-factor}. The possibility of constraining
fractal space dimensionality from astrophysics and other areas \citep{Caruso e Oguri},
also points to a fractal dimension very close to an integer value,
but still noninteger. Thus, the importance of this formalism that
it may indicate some higher-order effects that might exist in the
systems under study, without the cumbersome calculations of QED.

Other interesting applications are referred in the result of works
from, for example, Ostoja et al., see e.g. Refs.\citep{Ostoja,Ostoja-Waves}
with continuum models on fractal porous domains. These works use the
Jumarie's definition of derivatives and are highly cited, in a positive
way, but one of the main criticisms of the approach, as it can be
seen, for example, in Refs. \citep{Tarasov 2016-Acustic,Tarasov Anisotropic},
indicates that the approach into consideration, makes use of the MRL
derivative and does have applicability. But here we have shown that
it is an approach valid within a low-fractionality limit and it is
in essence a local formalism. 

\medskip{}

The authors wish to express their gratitude to FAPERJ-Rio de Janeiro
and CNPq-Brazil for the partial financial support.

\textbf{\bigskip{}
}

\end{document}